\documentclass[sigconf]{acmart}
\AtBeginDocument{%
  \providecommand\BibTeX{{%
    \normalfont B\kern-0.5em{\scshape i\kern-0.25em b}\kern-0.8em\TeX}}}

\usepackage{makecell}
\usepackage{multirow}
\usepackage{natbib}
\usepackage{colortbl}
\usepackage{multirow} 
\usepackage{hyperref} 
\usepackage{enumitem}
\usepackage[utf8]{inputenc}

\copyrightyear{2024} 
\acmYear{2024} 
\setcopyright{rightsretained} 
\acmConference[CHI '24]{Proceedings of the CHI Conference on Human Factors in Computing Systems}{May 11--16, 2024}{Honolulu, HI, USA}
\acmBooktitle{Proceedings of the CHI Conference on Human Factors in Computing Systems (CHI '24), May 11--16, 2024, Honolulu, HI, USA}
\acmDOI{10.1145/3613904.3641964}
\acmISBN{979-8-4007-0330-0/24/05}

\begin{document}

\title[ChatGPT and Young Professionals' Perception of Productivity and Accomplishment]{``If the Machine Is As Good As Me, Then What Use Am I?'' -- How the Use of ChatGPT Changes Young Professionals' Perception of Productivity and Accomplishment}


\author{Charlotte Kobiella}
\affiliation{
    \institution{Center for Digital Technology and Management}
    \city{Munich}
    \country{Germany}}
\email{kobiella@cdtm.de}
\additionalaffiliation{%
  \institution{LMU Munich}
  \city{Munich}
  \country{Germany}}

\author{Yarhy Said Flores López}
\affiliation{%
  \institution{Center for Digital Technology and Management}
  \city{Munich}
  \country{Germany}}
\email{yarhy.flores@cdtm.de}
\additionalaffiliation{%
  \institution{Technical University of Munich}
  \city{Munich}
  \country{Germany}}

\author{Franz Waltenberger}
\affiliation{
    \institution{Center for Digital Technology and Management}
    \city{Munich}
    \country{Germany}}
\email{waltenberger@cdtm.de}
\additionalaffiliation{%
  \institution{Technical University of Munich}
  \city{Munich}
  \country{Germany}}

\author{Fiona Draxler}
\affiliation{%
  \institution{University of Mannheim}
  \city{Mannheim}
  \country{Germany}}
\additionalaffiliation{%
  \institution{LMU Munich}
  \city{Munich}
  \country{Germany}}
\email{fiona.draxler@uni-mannheim.de}

\author{Albrecht Schmidt}
\affiliation{%
  \institution{LMU Munich}
  \city{Munich}
  \country{Germany}}
\email{albrecht.schmidt@ifi.lmu.de}

\renewcommand{\shortauthors}{Kobiella et al.}


\begin{abstract}
    Large language models (LLMs) like ChatGPT have been widely adopted in work contexts. We explore the impact of ChatGPT on young professionals' perception of productivity and sense of accomplishment. We collected LLMs' main use cases in knowledge work through a preliminary study, which served as the basis for a two-week diary study with 21 young professionals reflecting on their ChatGPT use. Findings indicate that ChatGPT enhanced some participants' perceptions of productivity and accomplishment by enabling greater creative output and satisfaction from efficient tool utilization. Others experienced decreased perceived productivity and accomplishment, driven by a diminished sense of ownership, perceived lack of challenge, and mediocre results. We found that the suitability of task delegation to ChatGPT varies strongly depending on the task nature. It's especially suitable for comprehending broad subject domains, generating creative solutions, and uncovering new information. It's less suitable for research tasks due to hallucinations, which necessitate extensive validation.
\end{abstract}




\begin{comment}
\begin{teaserfigure}
  \includegraphics[width=\textwidth]{sampleteaser}
  \caption{Seattle Mariners at Spring Training, 2010.}
  \Description{Enjoying the baseball game from the third-base
  seats. Ichiro Suzuki preparing to bat.}
  \label{fig:teaser}
\end{teaserfigure}
\end{comment}


\maketitle

\section{Introduction}

Using tools to simplify tasks has been closely linked to human development for the last 10,000 years. Tools have become more sophisticated and have enabled us to automate many manual tasks in the workplace. With digital AI-based tools, opportunities for automation have become widely available, including for cognitive and creative tasks. Besides curiosity, the desire to be more efficient and make work easier explains the interest in ChatGPT and similar services. The current adoption rate suggests that these new tools will revolutionize knowledge work through automation and impact white-collar workers across various fields. 
Since its release in November 2022\footnote{\url{https://openai.com/blog/chatgpt} (last accessed: 09/09/2023)}, ChatGPT has seen rapid adoption with more than an estimated 100 million monthly users just two months after launch \citep{chow_how_2023}.
Generative artificial intelligence (GenAI) marks a groundbreaking shift in AI technology as it enables machines to imitate human creativity, greatly enhancing its utility in creative and knowledge-intensive professions \cite{dwivedi_opinion_2023, berg_capturing_2023}.

Contrary to earlier technological shifts, scholars predict a significant effect of AI automation on knowledge work, particularly high-level tasks \citep{holford_future_2019}. 
Recent research echoes these findings and suggests that in the future, advanced language models will influence the tasks of around 80\% of workers, particularly those in knowledge-intensive fields \citep{eloundou_gpts_2023}. Yet previous research about the effects of automation in knowledge work has not been conclusive.
On the one hand, automation can lead to deskilling, displacement of workers, and an increase in unemployment \citep{acemoglu_automation_2019, bessen_automatic_2019, acemoglu_race_2018, susskind_model_2017}.
On the other hand, automation enhances the capabilities of human workers, increasing their productivity and wages \citep{boustan_automation_2022, kanazawa_ai_2022, agrawal_artificial_2019, hoffman_discretion_2018, kleinberg_human_2017}. 

ChatGPT has been widely adopted in different fields of knowledge work such as consumer science \citep{paul_chatgpt_2023}, education \citep{mollick_assigning_2023, kasneci_chatgpt_2023}, natural science \citep{madani_large_2023, ting_chatgpt_2023}, and software engineering \citep{ma_scope_2023}. Large language models (LLMs) and specifically ChatGPT are incorporated into knowledge work because they promise increased productivity \citep{peng_impact_2023, eloundou_gpts_2023, gilardi_chatgpt_2023, ritala_transforming_2023}. For example, a recent quantitative study by \citet{noy_experimental_2023} reports increased productivity and quality of work output when using ChatGPT for knowledge work. Nonetheless, several studies also show adverse effects of ChatGPT in the workplace. \citet{ahmad_impact_2023} report that GenAI promotes laziness and reduces critical thinking, and \citet{boyaci_human_2022} show that GenAI increases workloads in decision-making.

However, we know very little about the qualitative experience of professionals using LLMs, and notably, how the shift of ownership and individual contribution of work affects individuals' self-reported productivity and sense of accomplishment. In a world where LLMs can readily assist in and even automate specific tasks, there is a risk that knowledge workers may no longer find their work as meaningful as before, potentially leading to a diminished sense of accomplishment and decreased job satisfaction, motivation, and engagement. Following a preliminary study to find current use cases for LLMs and its drivers of adoption, we conducted a two-week diary study with 21 participants to investigate how young professionals reflect on their ChatGPT use and how the use of ChatGPT influences self-reported productivity and sense of accomplishment.

We find that knowledge workers agreed that their perceived productivity increased since they could perform more tasks in less time.
However, participants felt the output generated by ChatGPT was sub-par, and they felt a need to edit and post-process this output. 
Interestingly, this post-processing was also a driver of accomplishment since it was related to their sense of ownership and personal contribution to the work.  
We also elicit the participants' best practices for using ChatGPT.
As the lines between human and AI labor become increasingly blurred, it is essential to understand how these technologies affect workers on an individual level so that we can design human-AI interaction that is beneficial to the users.

The contributions of our paper are three-fold: Considering the example of ChatGPT, (1) our research extends the primarily quantitative understanding of the productivity effects of LLMs on knowledge workers and shows how participants preserve their sense of accomplishment; (2) we show use cases for which LLMs prove to be useful; (3) and provide a snapshot of best practices on how to use LLMs for enhanced productivity and sense of accomplishment. 
\section{Related Work}

In recent years, the emergence of LLMs and especially GenAI tools like ChatGPT have introduced a new era in AI systems. 
ChatGPT can engage in dynamic and contextually-driven conversations, emulating human-like conversational abilities, and its distinctive focus on natural language interaction facilitates more nuanced and dynamic exchanges between users and the system.
Notably, the intuitive interface and chatbot-like designs like ChatGPT and Google Bard are lowering the entry barrier for individuals and organizations seeking to leverage AI capabilities, providing GenAI tools with the potential to transform the work practices of various professions \citep{roose_how_2023, ghosh_genai_2023, menon_chatting_2023}.

GenAI tools have been used in the hope of bolstering the productivity of knowledge workers \citep{dwivedi_opinion_2023, berg_capturing_2023, biermann_tool_2022} and providing valuable support in problem-solving, idea generation, and conceptual development \citep{lee_coauthor_2022, di_fede_idea_2022, chung_talebrush_2022}, key facets of knowledge-intensive careers. 
Therefore, we investigate the subjective assessment of one's productivity within this study.

\subsection{Productivity in Knowledge Work}
Traditionally productivity is defined as the ratio of output versus input.
This metric could be easily measured for traditional factory workers as output was mostly tangible products, and input and output could be easily distinguished \citep{drucker_knowledge-worker_1999}.
However, assessing the productivity of knowledge workers is not as straightforward since the outputs they produce are specific to their respective fields and not easily measurable. The methods and resources they use vary widely among different knowledge workers. It is, therefore, difficult to establish a standardized measure for their input \citep{ramirez_measuring_2004}.
Moreover, \citet{kim_understanding_2019} and \citet{guillou_is_2020} show that knowledge workers do not base their productivity on quantifiable results but show how self-reported productivity is influenced by emotional measures such as emotional state and attitude towards work.
Hence, performative and measurable productivity does not necessarily align with knowledge workers' perceived productivity.

Recently \citet{noy_experimental_2023} evaluated the productivity of knowledge workers working with ChatGPT.
They show that using ChatGPT increased productivity, i.e., creating more output in less time with a higher quality of work results. 
Participants who showed higher results in a control task took even less time to complete the task than participants who showed lower results in the control task.
Quality improvements were more prominent for participants whose first unaided control task was graded lower.
However, we know little about the perceived productivity of knowledge workers when using ChatGPT.

In the era of automation, knowledge work has become more and more creative \citep{loo_creative_2017}, and the main value of human labor in the automation age is predicted to lie in creativity and social skills \citep{holford_future_2019,frey_future_2017, sokol_importance_2021}.
In pursuit of a competitive edge, companies are actively exploring avenues to fully harness the creative potential within their workforce \citep{ferguson_human_2010, george_when_2001}.
This underscores the paramount significance of creative capabilities in knowledge-intensive tasks.
These creative capabilities are rooted in the individuals' creative self-efficacy, and research shows that creative self-efficacy can significantly influence the creative performance in the work domain \citep{puente-diaz_creative_2015, carmeli_influence_2007, tierney_pygmalion_2004}.

\subsection{Self-Efficacy and Personal Accomplishment}

\textit{Creative self-efficacy} is one's confidence in the capacity to generate creative results \citep{tierney_creative_2002} with creativity as an act that yields innovative and valuable ideas, products, or performances \citep{sternberg_concept_1999}.
The concept of self-efficacy is rooted in the social cognitive theory by \citet{bandura_explanatory_1986}, where self-efficacy is a key element. 
Based on \citet{bandura_social_1999}, human motivation is driven by the belief in the capability to accomplish a task and the expectation regarding the outcomes of the actions.
Furthermore, self-efficacy significantly shapes an individual's motivation, emotional states, and behaviors, and these are influenced more strongly by personal beliefs rather than objective reality \citep{bandura_explanatory_1986}.
Experiences of mastery are the most influential source of self-efficacy and are sourced from personal accomplishments \citep{bandura_explanatory_1986}.
Recent literature also postulates a high correlation between personal accomplishment and \textit{creative} self-efficacy \citep{bang_personal_2016, karaboga_creativity_2022}.

\textit{Personal accomplishment} is connected to how competent and successful people feel in their work \citep{maslach_maslach_1997}.
\citet[p. 5]{maslach_measurement_1981} define personal accomplishment as a ``feeling of competence and successful achievement in one's work with people''.
If people perform a job, the feeling of accomplishment gives them confidence and assures them that they can successfully do similar tasks. Creative roles are inherently demanding, often requiring tailored approaches to address the challenges that arise \citep{unsworth_creative_2005}.
However, a history of successfully mastering tasks can empower individuals to address problems and develop innovative solutions \citep{hsu_creative_2011}. 
Therefore, an individual who has experienced significant personal accomplishment in their work is likely to feel confident in their ability to handle creative tasks, even those more challenging than they have faced before. 
This emphasizes the close connection between personal accomplishment and the development of both skill and the mastery of tasks \citep{elliot_test_1999}, and this connection directly impacts one's self-efficacy beliefs about their creative capabilities.
According to \citet{bandura_self-efficacy_1997}, individuals' perceptions of their accomplishments, formed from previous successful experiences, significantly contribute to enhancing their efficacy beliefs.

Workers with a strong belief in their creative abilities, i.e., workers with a higher creative self-efficacy, tend to have higher confidence in their job performance through their self-assessed personal accomplishment \citep{tierney_creative_2002}. 
As personal accomplishment increases, so does an individual's motivation, productivity, creativity, and problem-solving abilities \citep{hsu_creative_2011}.
The boost in creativity, in turn, improves task performance, which again leads to higher productivity \citep{karaboga_creativity_2022}. 

We are now at a pivotal point where humans still know how to perform knowledge work tasks independently and still feel capable since ChatGPT was only recently launched. 
\citet{noy_experimental_2023} show that ChatGPT, for now, does not influence individual self-efficacy.
Thus, in our research, we instead want to focus on ChatGPT's impact on personal accomplishment.
Now that knowledge work becomes increasingly dependent on creative attributes of work and technology like ChatGPT being able to even automate creative tasks, we need to understand the effects on personal accomplishment to evaluate the current implications of LLMs on knowledge work.

Our research aims to understand how knowledge workers currently use LLMs such as ChatGPT, how they assess their productivity on a quality and quantity level, and how using ChatGPT influences their personal accomplishments. 
Our study will also elicit a snapshot of best practices for using ChatGPT in knowledge work based on the current adoption.
\section{Methods Overview}
\label{sec:method}

We investigated productivity and personal accomplishment with LLMs in a two-phase study. First, we gathered usage scenarios and refined our research focus in a preliminary study (pre-study). Second, the context of the use cases identified during the pre-study served as the basis for the subsequent two-week diary study. We administered daily surveys to explore and understand how ChatGPT influenced knowledge workers' sense of productivity and accomplishment in their daily tasks.
 
For both the pre-study and the diary study, our sample consisted of young professionals enrolled in a master's level educational program focusing on entrepreneurship and innovation.
Participants of the pre-study were recruited from our local network at the educational institution. 
For the diary study, we recruited the participants from a pool of students enrolled in a seven-week full-time course at the institution focusing on trend research and ideation.
This population is interesting for our studies as they represent early adopters of the technology and are likely to have already used LLMs such as ChatGPT in their academic and professional careers. Furthermore, this procedure allowed us to control for their level of work experience and ensure comparability across participants.

During the pre-study, we identified relevant use cases of LLMs in the context of knowledge work.
Building upon the outcomes of the pre-study, we conducted a two-week diary study with 23 participants, of whom 21 completed the study. 
This phase allowed us to further explore and understand how ChatGPT influenced knowledge workers' sense of productivity and accomplishment in their daily tasks.

\begin{figure*}
    \centering
    \includegraphics[width=\linewidth]{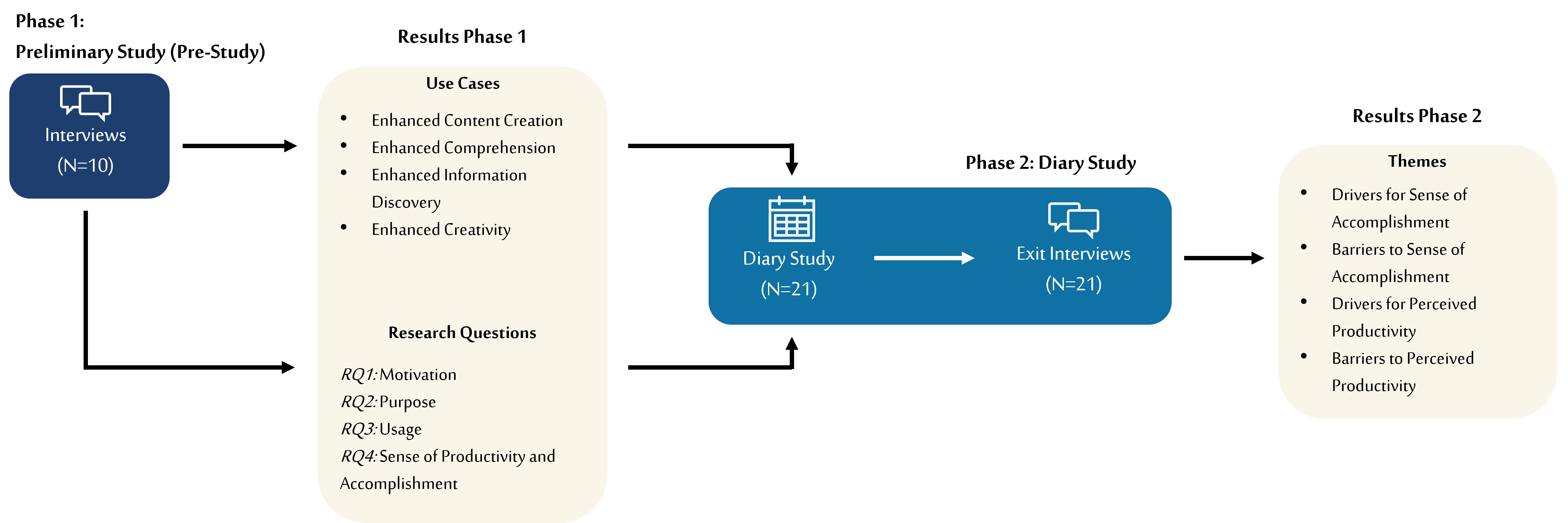}
    \caption{The methodology overview outlines the progression from initial use cases and research questions to the themes identified from the results of our diary study.}
    \label{Meth.Overview}
    \Description{Flowchart labeled 'Methodology Overview' with four elements connected by action links. The start state on the left is 'Phase 1 Preliminary Study'. The 'Phase 1: Preliminary Study' leads to 'Results Phase 1', which is divided into use cases and research questions. The 'Results Phase 1' leads to 'Phase 2: Diary Study', which is divided into diary study and exit interviews. The end state is 'Results Phase 2'.}
\end{figure*}

Figure \ref{Meth.Overview} illustrates the study design and the progression from the pre-study to the diary study.

\section{Phase 1: Pre-study Methods \& Results} \label{Pre-Study}

The core objective of the pre-study was to gain a comprehensive understanding of the specific contexts in which LLMs provide the most substantial value, both within the academic sphere and in professional contexts.
Additionally, we sought not only to explore the user experience (UX) challenges encountered by participants during their interactions with ChatGPT but also to understand the underlying reasons for these challenges.

\subsection{Methods}
To gain deeper insights into knowledge workers' LLM usage contexts, we employed a data collection method centered around semi-structured interviews, a well-established approach commonly utilized in qualitative research \citep{gill_methods_2008}.
These interviews allowed us to engage with participants and explore the intricacies of their experiences and perceptions regarding LLMs. For the pre-study, our sample consisted of ten young professionals with diverse work experiences, including engineering, business development, consulting, legal, and human resources.
We evaluated the interviews using an affinity diagram as it is a suitable qualitative analysis method to organize and synthesize data from semi-structured interviews \citep{dam_affinity_2022}. This method allowed us to refine our research questions and identify use cases for the subsequent diary study.
We concluded the pre-study after ten interviews when no new use cases emerged.
Other studies conducted with a homogeneous group of participants have shown similar sample sizes when reaching saturation \citep{hennink_sample_2022}.

\subsection{Results}

The pre-study revealed diverse uses of LLMs within the context of knowledge work, primarily for content creation, information discovery, creativity, and comprehension:

\begin{enumerate}
\item \textbf{Enhanced Content Creation:} ChatGPT serves as an integral tool for efficient content creation and refinement, aiding users in enhancing text structure and coherence.
\item \textbf{Enhanced Information Discovery:} ChatGPT accelerates the process of targeted information retrieval, eliminating the need to traverse multiple, broader sources, thereby granting users a head-start in topic exploration.
\item \textbf{Enhanced Creativity:} ChatGPT fosters rapid ideation, positioning it as a potential co-creative ally.
\item \textbf{Enhanced Comprehension:} ChatGPT simplifies complex concepts, facilitating rapid understanding through relatable explanations and providing an environment for deeper, paced exploration.
\end{enumerate}

Despite its utility, participants encountered UX challenges such as text length limitations, tone setting, interaction constraints, and navigation issues within the chat interface. Notably, while some participants reported enhanced productivity, opinions varied regarding the impact on the quality of work. 

These insights, reflecting both the utility and limitations of ChatGPT in professional settings, led to the formulation of four distinct research questions for further investigation in the subsequent diary study phase:

\begin{enumerate}
\item[\textbf{RQ1}] Motivation: What motivates knowledge workers to incorporate ChatGPT into their workflows?
\item[\textbf{RQ2}] Purpose: Which knowledge-centric tasks are most frequently delegated to or augmented by ChatGPT?
\item[\textbf{RQ3}] Usage: How do knowledge workers engage with ChatGPT?
\item[\textbf{RQ4}] Sense of Productivity and Accomplishment: How does the integration of ChatGPT in the workflow affect the individuals' sense of productivity and accomplishment?
\end{enumerate}

The outcomes of our pre-study indicated that productivity emerged as a predominant motivation among participants for using ChatGPT. 
This insight was particularly instrumental in formulating \textbf{RQ1}. 
Recognizing the emphasis on productivity, we decided to further investigate whether this motivation would consistently appear as a primary driver in a larger sample over a longer period of time.

The evaluation of the pre-study interviews also yielded distinct use cases, highlighting the varied purposes for which ChatGPT is employed within the context of knowledge work. 
The diverse applications reported by participants in the pre-study were instrumental in formulating \textbf{RQ2}. 
This question led us to investigate if a broader study would reveal a predominant theme in the usage of ChatGPT or uncover new, previously unidentified applications.

Similarly, we identified distinct interaction patterns with ChatGPT, such as developing effective prompts and refining the tool's outputs. This finding shaped \textbf{RQ3}, guiding us to determine whether these observed patterns with ChatGPT are consistent in a broader sample or if new engagement strategies emerge, thereby deepening our understanding of ChatGPT's role in knowledge work.

While the aspect of productivity and accomplishment forms the core of our entire study, it is inherently justified by the need to understand the overall impact of ChatGPT on work efficiency and satisfaction. This necessity laid the foundation for \textbf{RQ4}. Exploring productivity and accomplishment is central to comprehending the full scope of ChatGPT's integration into professional workflows.
\section{Phase 2: Diary Study Methods} \label{Diary Study}

Building on the foundational insights from the pre-study, we sought to further explore the detailed interactions of knowledge workers with ChatGPT. To understand how interactions with ChatGPT influence knowledge workers' sense of productivity and accomplishment, we conducted a two-week diary study, which allowed us to capture interactions and the subsequent feelings they evoke without losing contextual details \citep{bolger_diary_2003}.

\subsection{Participants}
For the diary study, we recruited 23 knowledge workers as outlined in \autoref{sec:method}. 
Participants who had already participated in the pre-study were excluded from the diary study. 
We instructed the participants to keep logs for at least seven (out of ten) weekdays to complete the study.
21 out of 23 participants fulfilled this condition.

The participants' ages ($ n=21 $) ranged from 20 to 27, with a mean age of 22.8 years ($ SD = 1.72$ ). Concerning gender distribution, 39.13\% identified as female, while 60.86\% identified as male. This demographic mirrors the age and gender distribution of OpenAI's users as reported by \citet{brandl_chatgpt_2023}. Specifically, 62.52\% of OpenAI's site visitors are aged between 18 and 34, and the gender split is 34.32\% female and 65.68\% male. The nationality distribution among our participants included 13 Germans, with one participant each from Austria, the Czech Republic, Egypt, India, Italy, Poland, Taiwan, and Uzbekistan.

Regarding educational attainment among the participants, 10 had completed high school, 8 held a bachelor's degree, and the remaining 3 possessed a master's degree. 
As outlined in \autoref{sec:method} all participants where part of the same master's level educational program at the time of our study. Students admitted to the program must have attained at least 10 weeks of full-time equivalent work experience in the field of technology and/or business, e.g., software engineering or consulting.
Of the 21 study participants, 9 had full-time work experience ranging from 13 to 24 months. Another 6 reported full-time work experience spanning 7 to 12 months. A further 5 had full-time work experience exceeding 24 months, while one had between 1 and 6 months of full-time work experience. The participants' academic backgrounds were Business and Economics (9), Computer Science (6), Natural Sciences (3), Engineering (1), Social Sciences (1), and Arts and Design (1).

\subsection{Study Setup \& Procedure} \label{sec:diary_study_setup}
In the diary study, participants were engaged in a seven-week full-time course divided into three distinct phases: the trend, scenario, and ideation phases.
The aim of the course is to develop a consulting-style industry report that provides a comprehensive overview of the future business environment by analyzing trends, describing future scenarios, and developing business ideas.
As the actual tasks of knowledge workers vary widely depending on the field they are working in, we focused our study on the creation of the industry-report as an example for knowledge work.

Our study was conducted during the second and third weeks of the course. 
In the first week of our diary study, participants were involved in initial tasks like desk research, interviews, fact-finding, and data validation, culminating in executive summaries and content editing.
As they transitioned into the scenario phase of the course and the second week of our diary study, their focus shifted to creative processes, including ideation and narrative building, alongside ongoing content refinement aimed at constructing detailed future scenarios based on the identified trends.
These tasks can be best mapped to learning, communication and project-related tasks which 
\citet{kim_understanding_2019} show to be among the most prevalent task categories of knowledge work. 

To gather daily information on the participants' use of ChatGPT, we decided on a time-based design \citep{bolger_diary_2003} and set up the diary study as a feedback diary study \citep{carter_when_2005} where participants reported their ChatGPT usage in a Google survey which we shared via a Slack group chat every day at 6:00 p.m. as this was usually the end of the participants' working day.
This approach allowed us to collect timely and consistent data regarding their interaction with ChatGPT, their productivity, and their sense of accomplishment while giving participants a chance to reflect on their day and the role of ChatGPT in their overall daily output.
The surveys were closed after 24 hours to minimize retrospection bias \citep{mitchell_temporal_1997}.
We aimed to understand the qualitative reasoning behind the participants' productivity and accomplishment assessments and the impact of ChatGPT. 
The survey thus used quantitative scales for the productivity and accomplishment assessments based on a 6-point Likert scale \citep{likert_technique_1932} and free text fields to elaborate on the ratings.
We included two questions for participants who did not use ChatGPT on the respective day, capturing their rationale for not using ChatGPT.
An overview of the entire survey can be found in appendix \ref{diary_study_survey}.
Participants received a EUR 20 remuneration upon completion of the study.

Before starting our diary study, we instructed the participants in an hour-long introduction session.
In this session, we explained the study set-up and goal and made sure that everyone had an OpenAI\footnote{\url{https://platform.openai.com/apps} (last accessed: 09/09/2023)} account to access ChatGPT.
Participants could use their existing OpenAI accounts, but we instructed them to only use the freely available version of ChatGPT based on the GPT-3.5 model (3 August 2023 version) to ensure the comparability of the results.
In the introduction session, we also shared best practices on prompting to ensure that all participants could use ChatGPT effectively.

Upon completion of the diary study, we examined the open-ended responses in the free text fields to better understand how participants justified their accomplishment and productivity evaluations. 
Two authors individually analyzed a portion of the entries (139 items; 77.22\%) employing the thematic analysis method as described by \citet{braun_using_2006} as it offers a theoretically adaptable approach that allows for the identification and examination of themes or patterns within qualitative data.
The survey responses were analyzed using an inductive, open-coding approach \citep{braun_using_2006, nowell_thematic_2017} with a focus on identifying themes related to factors influencing the assessment of accomplishment and productivity. 
Some free text fields were multi-coded if participants included different rationales for their assessment. 
For the keyword tagging and theme building, we used the qualitative research software Condens\footnote{\url{https://www.condens.io/} (last accessed: 12/11/2023)}.
After agreeing on the coding approach and settling any differences in coding through several discussions, one author coded the remaining data.
Our analysis resulted in 42 different keyword tags and 719 tagged observations.

After coding the daily forms, we conducted individual online exit interviews with each participant. 
Our goal for the exit interviews was to contextualize the results of the daily forms and probe deeper into the identified themes and emergent patterns.
We used a semi-structured interview guide to capture nuances in their survey answers (see appendix \ref{exit_interviews_guide} for the interview guide).
To facilitate subsequent data analysis, all interviews were recorded and transcribed using the AI software tool Airgram\footnote{\url{https://www.airgram.io/} (last accessed: 09/09/2023)}.
\section{Results of the Diary Study} \label{Results}

Participants in our pre-study revealed their motivation for using ChatGPT and for what tasks they used ChatGPT (cf. \autoref{Pre-Study}). 
In the following section, we report on the findings of our diary study, which aims to further elicit the distinct impact of incorporating ChatGPT into a knowledge worker's workflow and its impact on perceived accomplishment and productivity.
Through thematic analysis, we identified three central themes – \textit{Sense of Ownership}, \textit{Smart Use of ChatGPT,} and \textit{Task Completion} – each contributing uniquely to the participants' sense of accomplishment. 
Additionally, we discovered that while ChatGPT significantly enhanced productivity through aspects like \textit{Time Efficiency}, \textit{Increased Output}, and \textit{Streamlining Information Gathering}, it also presented challenges. 
Although participants reported getting tasks done more efficiently, the output provided by ChatGPT did not always meet their quality standards.
These included issues of \textit{Limited Reliability}, \textit{Grammar and Spelling Issues}, and \textit{Generic Output}, underscoring the need for post-processing and critical reflection on AI-generated content.

\begin{figure}[h]
\centering
\includegraphics[width=\columnwidth]{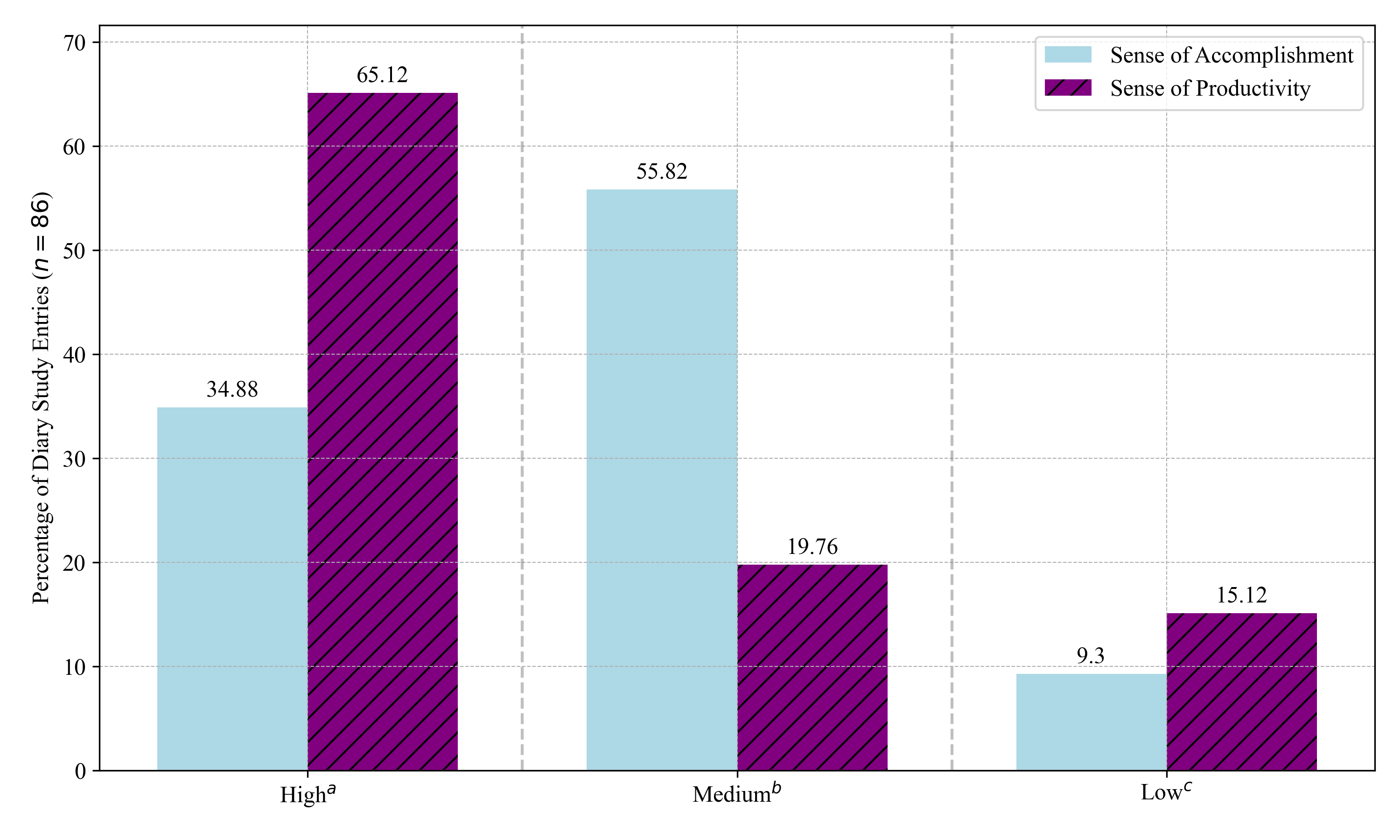}
    \begin{minipage}{8cm}
        \vspace{1.0em} 
        \textsuperscript{a}High: Experience rated at 5 or 6 on a 6-point Likert scale. \\
        \textsuperscript{b}Medium: Experience rated at 3 or 4 on a 6-point Likert scale. \\
        \textsuperscript{c}Low: Experience rated at 1 or 2 on a 6-point Likert scale. \\
        \hrule
        \vspace{0.3em} 
    \end{minipage}
    \caption{Likert scale ratings of diary study entries, reflecting varied levels of accomplishment and productivity as self-reported by participants. The chart quantifies participants' experiences, delineating them into high, medium, and low categories for each sentiment.}
    \label{fig:summary_statistics}
    \Description{This bar chart presents the summary statistics of the diary study entries, categorized by levels of Sense of Accomplishment and Sense of Productivity as reported by participants. Two sets of bars represent the percentage of diary study entries corresponding to each category. The left blue bars indicate the Sense of Accomplishment, with 'High' rated experiences at 34.88\%, 'Medium' at 55.82\%, and 'Low' at 9.30\%. Adjacent purple-striped bars denote Sense of Productivity, showing 'High' experiences at 65.12\%, 'Medium' at 19.76\%, and 'Low' at 15.12\%. Ratings are based on a 6-point Likert scale, with 'High' being 5 or 6, 'Medium' 3 or 4, and 'Low' 1 or 2. The chart illustrates a higher prevalence of productivity in high-rated experiences compared to the sense of accomplishment across the same levels.}
\end{figure}

During our two-week study period, we collected 182 ChatGPT Usage Journal entries recorded by 21 participants. Of these entries, 86 (47.25\%) indicated that participants engaged with ChatGPT on the respective day. 
Examining these interactions more closely, we found that 30 entries (or 34.88\% of the aforementioned 86) conveyed a sense of accomplishment, with participants rating their experience at either 5 or 6 on a 6-point Likert scale. Regarding productivity, 56 entries (equivalent to 65.12\% of those who used ChatGPT) expressed feeling productive by selecting a score of 5 or 6 on the same scale. A summary of the statistics for entries that utilized ChatGPT is provided in Figure~\ref{fig:summary_statistics}.

Upon conducting the thematic analysis of the gathered data, we found four central themes encompassing 15 subthemes. These themes and subthemes appear intrinsically linked to the users' sense of accomplishment and productivity when interacting with ChatGPT. A comprehensive overview of these themes, their respective subthemes, and illustrative quotes from the diary entries can be found in Table~\ref{ThemesAndSubthemes for Sense of Accomplishment and Productivity}.

In the following subsections, we will discuss each theme in detail, explain the subthemes, and share insights based on the diary entries and semi-structured exit interviews.

\subsection {Drivers for Sense of Accomplishment} \label{Drivers Sense of Accomplishment}
This theme encompasses three subthemes, each pinpointing a key factor contributing to the sense of accomplishment knowledge workers feel when using ChatGPT. The subthemes that fall under this theme are as follows:

\begin{enumerate}
    \item \textbf{Sense of Ownership:} Participants underscored that, although ChatGPT assisted in their tasks, their work's essence and core content originated from their own efforts. ChatGPT was perceived not as a replacement but as an enhancement tool, aiding them in presenting their intellectual property in a more refined and accessible format.
    This feeling of ownership is closely linked to their increased sense of accomplishment, as they perceived their original ideas and efforts as central to the task and value creation process, with ChatGPT serving primarily as a supportive collaborator in their creative journey. For instance, P8 noted in their diary entry, \textit{``I got everything done in time, and I also researched things independently to find sources for the statements from ChatGPT''}. Similarly, P6 expressed, \textit{``The quintessence and the content was still mine, but it was just packaged nicer. So, in a way, it just made my ``intellectual property'' better accessible''}.
    \item \textbf{Smart Use of ChatGPT:} Participants felt that employing ChatGPT strategically and efficiently contributed to their sense of accomplishment. Their comments highlighted instances where they employed the technology judiciously, achieving positive outcomes swiftly and with minimal back-and-forth interactions. As P14 remarked, \textit{``I was writing a few different prompts, and ChatGPT didn't do what I wanted. And then, with the fourth prompt iteration, it gave me exactly the right script. And we didn't even change anything. And that also makes me feel quite accomplished because our work was basically finding the right prompt''}. By making informed decisions on when and how to integrate ChatGPT into their tasks, participants could streamline their workflow and achieve effective results. P17 also noted, \textit{``I just found a shortcut...I can save so much time because I feel like you can save time with ChatGPT as you don't overcomplicate it''}. 
    \item \textbf{Task Completion:} Participants noted that ChatGPT played a pivotal role in task completion, contributing to their sense of accomplishment. Their comments underscored instances where they efficiently obtained the necessary information or assistance through ChatGPT, allowing them to make substantial progress on their assignments. P8 explained, \textit{``I understood things faster, going through Google and Python documentation didn't get me to the info I needed''}. By relying on ChatGPT's capabilities to provide insights, answers, or content, participants found themselves able to move forward with tasks they might have otherwise struggled with. This heightened efficiency in task completion instilled a sense of accomplishment as participants appreciated their tangible progress, often remarking on the satisfaction of completing their work effectively and efficiently. P18 also expressed, \textit{``I don't care who came up with the cool idea, the key thing is to have it''}.
\end{enumerate}

\subsection{Barriers to Sense of Accomplishment} \label{Barriers Sense of Accomplishment}
This subtheme captures moments when participants experienced diminished accomplishment while engaging with ChatGPT. Several factors contributed to this sentiment, such as the simplicity of the tasks at hand, difficulties in effectively prompting ChatGPT, dissatisfaction with the content quality, a decreased sense of ownership over the finished product, and occasional feelings of inferiority. The subthemes encompassed by this theme include:

\begin{enumerate}
    \item \textbf{Lack of Challenge:} Participants in the study highlighted instances where tasks aided by ChatGPT lacked the difficulty level they typically associated with a sense of accomplishment. As P3 remarked, \textit{``I feel accomplished due to my progress, but prompting required so little work that it doesn't feel like I worked enough. [...] It did not feel like an accomplishment to ask questions and get answers''}. Some expressed that the ease of obtaining responses or generating content with minimal effort diminished the perceived level of accomplishment. As P7 stated, \textit{``I don't like using ChatGPT for work, sometimes it feels like cheating''}. This subtheme underscores how, in certain scenarios, the simplicity and efficiency of using ChatGPT can inadvertently lead to a decreased sense of accomplishment.
    \item \textbf{Prompting Difficulties:} Participants encountered challenges when formulating prompts for ChatGPT that would yield the desired outcomes. Their frustrations became evident when ChatGPT failed to comprehend and respond effectively to their prompts. This subtheme underscores the interplay between participants' expectations and ChatGPT's responsiveness, revealing a source of diminished accomplishment. Participants highlighted their desire for clear and precise interactions with ChatGPT to achieve optimal results. The difficulties they faced in prompting ChatGPT to meet their expectations added a layer of complexity to their tasks, impacting their sense of accomplishment. As stated by P7, \textit{``I think I should learn how to prompt better and be more efficient with my use''}. P9 remarked, \textit{``I felt a sense of accomplishment but annoyed a couple of times as well when ChatGPT does not realize my prompts as I would like it to do''}. By exploring this subtheme, we gain insights into the critical role of effective communication with AI systems in shaping the user experience and their ultimate sense of accomplishment.
    \item \textbf{Quality Dissatisfaction:} Participants in the study expressed a sense of diminished accomplishment when they encountered issues related to the quality of ChatGPT's outputs. Their comments revealed a common concern: the need for higher-quality results that aligned with their expectations and requirements. Instances of dissatisfaction emerged when participants perceived the outputs as generic, rushed, or superficial, lacking the depth and precision they desired. P14 stated, \textit{``You can generate a lot of results, but they lack quality if you don't have the time to dig deeper yourself''}. The time constraints for refining ChatGPT's responses also played a role, leading to a compromise in output quality. P10 described this as \textit{``I think I would have probably done a bit more qualitatively higher work, but I would be less productive. So I always had to calculate a trade-off''}. Participants noted that while ChatGPT could generate numerous results, the overall quality often left room for improvement. This subtheme sheds light on the pivotal role of output quality in shaping participants' sense of accomplishment, emphasizing the importance of aligning AI-generated content with user expectations.
    \item \textbf{Diminished Sense of Ownership:} A reduced sense of ownership emerged among participants when they felt they had little control over the output or a diminished contribution due to ChatGPT's predominant role. Participants expressed that their contributions seemed less significant when ChatGPT played a more central role in their tasks. They felt that the work was more of ChatGPT's creation than theirs, distancing them from the creative process. This reduced attachment and ownership often resulted in a decreased sense of accomplishment. 
    P11 reflected on their feelings of detachment, stating, \textit{``Hm - on the one hand, I delivered a high-quality work - on the other hand: It was not ``my'' work, but ChatGPT's work''}.  
    \item \textbf{Inferiority:} Participants experienced a sense of inferiority when they perceived their creativity and contributions as overshadowed by ChatGPT's output. The comments reflected a feeling of inadequacy, where participants believed their ideas couldn't compete with the AI-generated content. \textit{``If the machine is as good as me, then what use am I?''} [P7]. Despite recognizing that the outcome depended on how they utilized ChatGPT's output, this sense of inferiority was linked to diminished feelings of accomplishment. This subtheme underscores the psychological impact of AI assistance on individuals' self-perception and sense of accomplishment related to their creative abilities.
\end{enumerate}

\subsection{Drivers for Perceived Productivity}

The impact of ChatGPT on the sense of productivity becomes evident in the four subthemes that emerged. These subthemes underscore how ChatGPT empowers participants to save time, increase their output, strategically outsource tasks, and streamline information gathering. Collectively, these aspects boost productivity and enhance participants' sense of accomplishment as they efficiently achieve their objectives and generate more output in the same timeframe.

\begin{enumerate}
    \item \textbf{Time Efficiency:} Participants noted that ChatGPT's ability to generate ideas, responses, and content rapidly outpaced their capacity for ideation and production. This time-saving aspect of ChatGPT was particularly valuable in meeting deadlines and completing tasks promptly. P8 remarked, \textit{``I'm not sure I could've gotten the things done before our deadline without ChatGPT''}. P19 made a similar statement, \textit{``We were short on time, and I would not have managed to be so extensive otherwise''}. It enabled participants to achieve their goals efficiently, sometimes eliminating bottlenecks in their workflow related to slow content generation or idea development. The accelerated text production and idea generation offered by ChatGPT contributed to the participants' heightened sense of productivity and accomplishment.
    \item \textbf{Increased Output:} Participants reported that ChatGPT enabled them to generate a larger volume of content, ideas, or questions than they could achieve independently. This amplified output capacity was especially beneficial for tasks requiring extensive content creation or creativity. \textit{``Research would have taken me a long time. I can spend less time understanding the topic and more time working on deliverables''} [P6]. By leveraging ChatGPT's capabilities to expand their creative or informational output, participants experienced a sense of productivity and accomplishment. They could accomplish more within a given timeframe, enhancing their overall task efficiency and effectiveness.
    \item \textbf{Outsourcing:} Participants remarked that ChatGPT could efficiently handle certain portions of their tasks, allowing them to allocate more time and effort to other aspects or tasks that required their unique expertise. For example, participants would let ChatGPT condense their written content while using the time to prepare for discussions. \textit{``Feels good to outsource this [kind of] work to ChatGPT because I don't enjoy it too much''} [P20]. This subtheme highlights the participants' ability to enhance productivity by strategically leveraging ChatGPT as a resource for specific task components, contributing to a sense of accomplishment through optimized work allocation and efficiency. \textit{``At this very particular task, I wanted to have the most efficient output possible''} [P6].
   \item \textbf{Lowering Entry Barriers for Information Gathering:} This subtheme highlights how ChatGPT catalyzes efficient information gathering and synthesis. Participants expressed that ChatGPT significantly reduced the initial stages of collecting data, insights, and information required for their tasks. By streamlining this process, ChatGPT enabled them to concentrate on more advanced aspects of their work, propelling them from the early stages to a more substantial portion of their tasks. \textit{``With ChatGPT, I save the whole 0 percent to 40 percent part and can fully concentrate on bringing it from 40 percent to 100 percent''} [P14]. This subtheme underscores how ChatGPT's ability to rapidly provide information contributes to increased productivity and a sense of accomplishment among participants.
\end{enumerate}

\subsection{Barriers to Perceived Productivity}

In our study, we could also identify barriers to productivity.
Although individuals primarily employ AI tools to augment their productivity rather than hinder it, participants in our study reported certain drawbacks to current GenAI solutions.
Through our thematic analysis of the collected data, we identified three distinct subthemes. These subthemes collectively address the users' necessity to refine and adjust the output provided by ChatGPT, shedding light on the nuances of their interaction and reliance on the tool. The subthemes include concerns about ChatGPT's reliability, grammar, and spelling issues and the generation of generic content.

\begin{enumerate}
    \item \textbf{Limited Reliability:} Participants noted that ChatGPT's reliability varied depending on the specific topic or request. They expressed the need for post-processing because they considered ChatGPT an unreliable source for certain types of information. Quality discrepancies were recurring, prompting participants to validate ChatGPT-generated data through alternative sources. They emphasized that while ChatGPT could assist in various tasks, it was not a dependable solution for final, conclusive work products. This subtheme illuminates the participants' hesitation to fully rely on ChatGPT's output, necessitating post-processing to ensure accuracy and reliability in their work. P3 described, \textit{``Due to mediocre quality, I had to look up these things myself''}. Similarly, P9 expressed, \textit{``So I would say, you always need to post-process things''}.
    \item \textbf{Grammar and Spelling Issues:} While participants found ChatGPT helpful overall, they noted that the output was not consistently error-free regarding language usage and mechanics. Consequently, additional time and effort were required for post-processing to rectify these issues. P19 noted, \textit{``Overall helpful, but grammar and writing not always correct, so additional time to fix is needed''}. 
    \item \textbf{Generic Output:} Participants expressed dissatisfaction with ChatGPT's tendency to produce surface-level or basic information, particularly in contexts requiring in-depth research or exploration. They noted that ChatGPT's output often served as a starting point but failed to provide comprehensive insights or detailed analyses. Consequently, participants needed to explore the topic more independently or consult additional sources to enrich the content. P14 stated, \textit{``Not satisfied with today's output; very basic and especially for in-depth research, not suited due to the lack of sources''}. On a similar note, P12 mentioned, \textit{``Did not dive as deep into the topic as I would have otherwise''}.
\end{enumerate}

\subsection{The Interplay of Personal Accomplishment and Productivity}

Collectively, the subthemes for the barriers to perceived productivity can be grouped into a need for post-processing of GenAI output.
Interestingly, although participants reported the need for post-processing as a barrier to productivity in the surveys, they also mentioned in the exit interviews how this boosted their perceived personal accomplishment.

In the exit interviews, we probed deeper on this theme by asking how the use of ChatGPT influences the participants' personal accomplishment and how the participants would feel if they got the perfect ChatGPT output based on their prompt (see questions (2) and (5) in appendix \ref{exit_interviews_guide}).
Their answers revealed that their personal accomplishment was mainly attributed to feelings of ownership, control, and contribution.

P10 elicited, \textit{``I think in rare cases, I used exactly the output from ChatGPT. Then my personal accomplishment was not as strong because it was not my work.''}
P14 mentioned, \textit{``It depends on how much I am still involved in the process. [...] If I use ChatGPT and make it myself to 100\%, then I feel very accomplished, and then I think ChatGPT even boosts this accomplishment because I probably wouldn't be able to get to the 100\% that easily without the groundwork from ChatGPT. [...] I feel unaccomplished when I directly have to use it.''}
P17 stated, \textit{``Well, in the beginning, I do feel accomplished because I'm like, okay, it just took me, like, half an hour, and I produced three pages of text. But then I'm also like, okay, but basically, what did I do? I just wrote a question. And I then copied the text. I have all of this text now, but when I read through it, it's just ChatGPT, and you can tell that. So, I kind of have to redo it. So, this is where my personal accomplishment comes from.''}
P6 drew a more nuanced picture by saying, \textit{``I feel like it's also this part of getting used to it as well. If that's just how everybody does it. And if you feel like you can't add value to it, then that's just state-of-the-art technology, I guess. And if you're aware of that, then I guess it doesn't have to necessarily change your sense of accomplishment. Although, now thinking about it, my first impression is that I probably would feel less accomplished''} before concluding \textit{``In the end, it's still about what you do with the output, though. So that's where I get my sense of accomplishment from.''}

Participants' reflections indicate that their sense of personal accomplishment varies depending on their level of involvement with ChatGPT's output; direct use of ChatGPT's content often leads to feelings of reduced accomplishment, while integrating and building upon its suggestions can enhance their sense of accomplishment in the creative process.

\begin{table*}
\centering
\caption{Themes and Subthemes for Sense of Accomplishment and Productivity}
\label{ThemesAndSubthemes for Sense of Accomplishment and Productivity}
\renewcommand{\arraystretch}{0.9}
\begin{tabular}{|p{3.0cm}|p{3.0cm}|p{8.0cm}|}
\toprule
\rowcolor{gray!30} 
\textbf{Theme} & \textbf{Subtheme} & \textbf{Quote} \\
\midrule
\multirow{6}{=}{Drivers for Sense of Accomplishment} & Sense of\newline Ownership & \textit{``The quintessence and the content was still mine, but it was just packaged nicer. So, in a way, it just made my ``intellectual property better accessible.''} [P6] \\
\arrayrulecolor{gray}\cline{2-3}
& Smart Use of\newline ChatGPT & \textit{``I think we made smart use of the technology and achieved good results.''} [P15] \\
\cline{2-3}
& Task Completion & \textit{``I took the answer for granted and was able to complete this part of my work. Whether it was true, I don't know.''} [P8] \\
\arrayrulecolor{black}\midrule
\multirow{4}{=}{Barriers to Sense of Accomplishment} & Lack of Challenge & \textit{``I feel accomplished due to my progress, but prompting required so little work, that it doesn't feel like I worked enough.''} [P3] \\
\arrayrulecolor{gray}\cline{2-3}
\cline{2-3}
& Prompting Difficulties & \textit{``I felt a sense of accomplishment, but annoyed a couple of times as well when ChatGPT does not realize my prompts as I would like it to do.''} [P9] \\
\cline{2-3}
& Quality\newline Dissatisfaction & \textit{``You can generate a lot of results, but they lack quality if you don't have the time to dig deeper yourself.''} [P14] \\
\cline{2-3}
& Diminished Sense of\newline Ownership & \textit{``Hm - on the one hand, I delivered a high-quality work - on the other hand: It was not ``my'' work, but ChatGPT's work.''} [P11] \\
\cline{2-3}
& Inferiority & \textit{``So it's sometimes sad to see that your own creativity can not compete most of the time.''} [P6] \\
\arrayrulecolor{black}\midrule
\multirow{4}{=}{Drivers for Perceived Productivity} & Time Efficiency & \textit{``ChatGPT enabled us to come up with ideas super quickly, and that would not have been possible as fast without it.''} [P15] \\
\arrayrulecolor{gray}\cline{2-3}
& Increased Output & \textit{``I could generate a lot of questions.''} [P14] \\
\cline{2-3}
& Outsourcing & \textit{``It would give me a lot of time left to do other stuff.''} [P3] \\
\cline{2-3}
& Lowering Entry Barrier\newline for\newline Information Gathering & \textit{``I can have more output in less time; with ChatGPT I save the whole 0\% to 40\% part and can fully concentrate on bringing it from 40\% to 100\%.''} [P14] \\
\arrayrulecolor{black}\midrule
\multirow{4}{=}{Barriers to Perceived Productivity} & Limited Reliability & \textit{``Quality is different depending on the topic/request and the data has always to be validated through other sources.''} [P9] \\
\arrayrulecolor{gray}\cline{2-3}
& Grammar and\newline Spelling Issues & \textit{``Overall helpful, but grammar and writing not always correct, so additional time to fix it needed.''} [P19] \\
\cline{2-3}
& Generic Output & \textit{``In the last research session, ChatGPT did not convince me to dive deeper and not only scratch the surface; I already had the overview and wanted to dive deeper, so I did it by myself reading scientific papers.''} [P14] \\
\arrayrulecolor{black}
\bottomrule
\end{tabular}
\medskip
\end{table*}
\section{Discussion} 

Our interviews and diary study investigated how young professionals use GenAI tools such as ChatGPT for knowledge work and how these tools impact their perceived productivity and sense of accomplishment. 
The main motivation for using ChatGPT for our participants was to increase their productivity (RQ1).
Across all participants, four prevalent themes for using ChatGPT in knowledge work emerged: (1) comprehension, (2) creativity, (3) information discovery, and (4) content creation (RQ2).
Our participants revealed strategies to conserve their sense of accomplishment, which was highly intertwined with their sense of ownership (RQ3).
We found that using these tools positively impacts their perceived productivity and sense of accomplishment even when the LLM assumes a substantial role in task performance (RQ4). Here, we discuss the implications of using LLMs on productivity and personal accomplishment. We further elicit when and how to use LLMs to ensure productivity and personal achievement.

\subsection{Perceived Accomplishment Using LLMs}

In sections \ref{Drivers Sense of Accomplishment} and \ref{Barriers Sense of Accomplishment}, we evaluated the drivers and barriers of perceived accomplishment. 
In this section, we discuss how interacting with the LLMs, i.e., prompting and post-processing, impacted the perceived sense of accomplishment. 
 
\subsubsection{Successful Prompting Fosters Sense of Accomplishment}

Participants repeatedly mentioned that efficient prompting made them feel more accomplished since it made them feel competent and knowledgeable.
Further, some participants mentioned that they deem the work effort to reside in the act of prompting.
These findings help to contextualize findings from other studies.
\citet{noy_experimental_2023}, for example, report that the self-efficacy of their participants who engaged with ChatGPT slightly increased even though participants primarily used it as a substitute for their own effort.
If prompting increases participants' sense of accomplishment, making clever use of LLMs might not diminish the individuals' self-efficacy.

Furthermore, we also identified the need for post-processing and contributing to the final result as an influential factor in one's sense of accomplishment.
However, we note that the initial prompting and the post-processing are intertwined concepts, i.e., post-processing of the output could be done with prompt refinements but also ``offline'' without the help of an LLM.
Thus, writing the prompt could lead to an increased sense of accomplishment regardless of the need to post-process the generated output.
Further, the heightened sense of accomplishment could also arise from vicarious experiences, e.g., seeing others being deemed tech-savvy and intelligent when performing tasks.
\citet{bandura_explanatory_1986} postulates both feelings, whether immediate or vicarious, as another source for self-efficacy and subsequent personal accomplishment.
With some participants acknowledging that prompting will be just another skill in their toolbox for productive working, it would be interesting to observe how and if these feelings of accomplishment last beyond the novelty of LLMs.

\subsubsection{Automation Impacts Ownership}

Ownership and control were among the most prevalent themes in our study.
In the exit interviews, participants reported that the need for post-processing gave them a sense of ownership in the task execution and control over the outcome.
Whereas some research looks into AI as a human-like team member and companion (e.g., \citep{flathmann_modeling_2021, mallick_designing_2022}), participants in our study reported that they saw the LLM as an aid rather than a collaborator.
They felt their ideas and efforts were still at the forefront, with the LLM being a tool to use according to their needs.

This is an interesting finding as there are opposing results in computer-supported cooperative work (CSCW). 
For example, \citet{zhang_ideal_2021} show that in multiplayer online games, AI is perceived as a tool and that the AI is expected to bring distinct skills to the team.
By contrast, a study by \citet{wang_human-ai_2019} among data science practitioners postulates that AI is a ``first-class subject'' in data science rather than a tool and a study by \citet{hayashi_can_2017} implies that in courtroom decision-making the AI system may be considered more trustworthy than the human counterpart, hinting that different professions call for different ways of human interaction based on the skills and methods of working.
In other instances, where participants felt like ChatGPT's assistance overshadowed their creative input, the use of ChatGPT had a diminishing effect on their personal accomplishments.
The desire for a more significant personal role in tasks and creative processes influenced their satisfaction levels and perception of accomplishment.

This is in line with substantial research in psychology where it has been demonstrated that the higher the subjective involvement, the higher the sense of ownership (see \citet{pierce_state_2003} for an overview).
 Participants could not come up with an exact number when asked in the exit interviews what their threshold for assigning ownership to themselves was. Instead, they mentioned that their need for post-processing also stems from a need to have the ``last say'' and thus control the outcome.

Generally, the sense of control is a prominent theme in the research of AI-mediated communication (AI-MC) \citep{mieczkowski_examining_2022}.
AI-MC primarily centers on AI tools that act on a communicator's behalf to achieve relational or communications objectives \citep{hancock_ai-mediated_2020}, which can be mapped to the tasks participants undertook in the use case of content creation.
In recent AI-MC research, \citet{mieczkowski_examining_2022} demonstrated that participants in writing tasks aim to maintain a holistic feeling of agency where one pivotal dimension for the agency is the feeling of control over the process and outcome.
Whereas the notion of agency is also prominent in computational co-creativity, \citet{koch_agency_2021} suggest that the system's pro-activity and adaptability influence it.
Similarly, a study by \citet{oh_i_2018} found that when co-creating creative output with an AI, users preferred detailed instructions from the AI but wanted to make all decisions during tasks and often anthropomorphized the AI.
It would be interesting to investigate how the perception of agency differs for our four reported use cases.

\citet{draxler_ai_2023} also discuss ownership in the light of AI-aided text generation.
Their findings postulate an ``AI Ghostwriter Effect'': users claim authorship even if they don't feel like owners of the text.
They also show that the sense of ownership increases with the sense of control and leadership one has in task performance.
This is in line with our findings as participants in our study could decide freely which and how many parts of a task they would outsource to the AI.
Notably, in our study, users repeatedly mentioned that they still feel in control of the outcomes since they made the final touches.

Our results expand these findings and suggest that the individual's agency and personal accomplishment are affected by how they decide to engage with AI. 
The experiences shared by participants underscore the significance of maintaining a harmonious balance between AI support and individual input, suggesting that when AI takes a leading role, it can overshadow a person's sense of ownership and, by extension, their sense of accomplishment.

\subsection{Perceived Productivity Using LLMs}

The participants reported productivity as a subjective feeling of efficiency, i.e., being able to finish tasks faster. 
Further, participants also attributed their \textit{ability} to perform a task faster due to using the LLM as a driver for productivity.
We want to contextualize these findings with the hedonic/pragmatic model of user experience (see \citet{hassenzahl_hedonicpragmatic_2007} and the unified theory of acceptance and use of technology (UTAUT2) (see \citet{venkatesh_consumer_2012}).

\subsubsection{Optimising for Objective Productivity}

\citet{hassenzahl_hedonicpragmatic_2007} divides user experience into two dimensions: hedonic (non-utilitarian), i.e., the ``be-goals'' and pragmatic (instrumental), i.e., the ``do-goals'' (see also \citep{hassenzahl_being_2007}).
The hedonic dimension is divided into sub-dimensions of identification and stimulation, while the pragmatic dimension relates to usability and usefulness. 
In our study, the ``do-goals'' related to the four identified use cases for which users opt to use ChatGPT, i.e., comprehension, creativity, information discovery, and content creation.

Participants showed how these could be further enhanced when optimizing for objective productivity.
They desired a more expansive range of tasks for ChatGPT, such as support in crafting prompts, including asking clarifying questions and understanding user intentions. A significant interest was expressed in a voice-based interaction mode, coupled with updated post-2021 information sources, including legal and academic content. Feedback also emphasized the importance of natively integrated web browsing and document upload capabilities, emphasizing the accuracy and verification of provided information. Users sought a more versatile, reliable, and user-friendly interaction with ChatGPT.

\subsubsection{Optimising for Perceived Productivity}
In other instances, participants also reported ``be-goals,'' such as being intelligent, competent, feeling in control, and being able to use state-of-the-art tools as their motivation to use ChatGPT to achieve their productivity goals. For example, P13 mentioned using ChatGPT facilitated \textit{``working smarter instead of harder''} and P8 reflected, \textit{``I still feel I did good work, but smarter by saving time''}. P5 also mentioned, \textit{``I feel proud of my explanation skills''} when a prompt works on the first try. 
The results can also be mapped to the hedonic/pragmatic user experience model, as the evaluation of the hedonic dimension is based on assessing the user's needs. 
Humans look for hedonic experiences and aim to fulfill needs when undertaking tasks and activities \citep{hassenzahl_needs_2010}.
The needs voiced by our participants boil down to the needs of \textit{autonomy} and \textit{competence}, which are among the most basic human psychological needs \citep{ryan_self-determination_2000, sheldon_what_2001}.

The hedonic/pragmatic user experience model emphasizes the importance of considering both pragmatic and hedonic aspects in understanding and evaluating user experience.
Pragmatic and hedonic qualities contribute equally to the overall quality of a product but are perceived by the user as independent \citep{hassenzahl_hedonic_2000}.
Similarly, \citet{guillou_is_2020} found that when knowledge workers were asked to characterize the metric of ``Time Well Spent'', they also look to account for emotional and physical well-being when assessing their time at work. Not only was what they worked on important for the participants but also how they worked (e.g., being efficient) and how they felt (e.g., satisfied or achieved).

Whereas recent research focuses solely on the usability aspects of LLMs (see \citet{vaithilingam_expectation_2022} for an example in software engineering), our study suggests that ``be-goals'' play an essential role in the acceptance and attractiveness of LLMs and should be discovered further since both, hedonic and pragmatic, dimensions are necessary for successful long-term adoption \citep{venkatesh_consumer_2012}.

\subsection{Usage of LLMs}

During most days in our diary study, participants decided to use the LLM; they also reported instances when LLMs might not be the right tool and even hindered productivity and sense of accomplishment.
We here discuss the reported findings on when and when not to use LLMs.
We further present reported best practices by participants of our study based on the four identified themes: (1) comprehension, (2) creativity, (3) information discovery, and (4) content creation.

\subsubsection{When Not to Use LLMs}

Participants opted not to employ ChatGPT due to various factors. 
These factors include mistrust stemming from previous negative experiences with ChatGPT and scenarios where ChatGPT was deemed unsuitable for meeting coordination, situations where its efficiency was questioned, moments when human decision-making skills were deemed necessary, and contexts primarily focused on discussions and interviews that did not require AI-generated content.

Trust in AI automation has been extensively studied (for an overview, see \citep{schwartz_enhancing_2023}).
In our context, mistrust refers to participants' reservations and doubts about ChatGPT's reliability, particularly concerning complex or critical topics. 
Participants expressed concerns that ChatGPT might generate inaccurate or fabricated information, leading to a lack of trust in its responses. 
This mistrust stems from experiences where ChatGPT provided misleading or false information on subjects that participants were knowledgeable about.
Our study's findings on mistrust in ChatGPT echo those by \citet{paul_chatgpt_2023}, which also highlights skepticism towards AI in consumer interactions. This similarity underscores mistrust as a pervasive challenge in AI applications, both in professional settings and consumer behavior.

Participants also reported instances where they deemed ChatGPT inefficient for specific tasks. 
Participants indicated that there are instances where ChatGPT is not the most efficient tool for the job, leading them to opt for alternative methods.
These tasks were mainly related to research or software engineering tasks.
Participants mentioned that they deemed ChatGPT not the right tool because they were better at the task, e.g., finding eligible sources and coding. 
Whereas this might be true now, more powerful LLMs could shift the dynamics in this realm in the future.

When asked in the exit interviews about which tasks or work they would not delegate to ChatGPT, participants mentioned tasks involving human collaboration, such as discussion and decision-making.
This underscores the distinct roles and tasks where human decision-making skills precede AI-generated assistance.
In situations where the primary focus was on dialogue and interaction among team members, ChatGPT was deemed unnecessary.

\subsubsection{Insights on LLM Best Practices}

The exit interviews highlighted different strategies for using LLMs that led to increased productivity and a heightened sense of accomplishment.
We deem the reported strategies an interesting snapshot of current human-AI interaction and summarize these interactions as best practices below.
We note that as LLMs evolve, so might the interaction strategies of users. 

\begin{enumerate}
    \item \textbf{Comprehension:}
    Participants mentioned ChatGPT as support in learning and understanding as it could break down findings and explain thoughts and frameworks.
    Recent research in the education space also postulates LLMs as a ``key enabling technology'' for innovative educational technologies \citep{kasneci_chatgpt_2023}.
    In our study, ChatGPT also allowed the participants to probe deeper into constructs they did not yet understand, individualizing and facilitating their learning journey.
    This led to high self-reported productivity as it sped up their comprehension process, paving the way to task execution.
    Further, participants reported a high personal accomplishment as they understood LLMs as a means to their learning achievement.
    \item \textbf{Creativity:}
    Overall, participants deemed the output generated by ChatGPT creative, especially for topics where they lacked prior expertise.
    In the exit interviews, participants contextualized the creativity aspect, classifying the output by ChatGPT as ``good inspiration'' and ``impulses,'' hinting that ChatGPT's perceived creativity is not everlasting.
    Participants still consider using LLMs in creative tasks as productivity-enhancing as they can build up on the output and further discuss it in humane group settings.
    Participants felt an increased sense of accomplishment, especially for known topics such as event planning, as they could ``compete'' with the AI.
    \item \textbf{Information Discovery:} 
    Participants could see the qualities of the LLMs when used for information discovery, such as getting initial insights into an industry. The LLMs were usually used as a first step to get an initial understanding and overview of the topic and which topics to explore. Participants then decided to dive deeper in subsequent Google searches to deepen their understanding and find more concrete input. This increased their sense of accomplishment and perceived productivity as they could spend less time understanding the topics and more time on the actual work result, feeling accomplished quicker.
    \item \textbf{Content Creation:} 
    Participants highlighted two primary methods of content creation using the LLM: (1) initiating content from scratch and (2) refining their pre-written text. For the former, the LLM effectively combated ``writer's block,'' allowing participants to avoid confronting a blank page, and was recommended especially for non-sensitive topics like social media campaigns. 
    
    In the latter scenario, the LLM enhanced their writing, making it more concise and fluent, particularly beneficial for those drafting in a second language. Although the LLM offered improvements, participants adjusted the content to maintain their unique style.
    A study by \citet{biermann_tool_2022} reveals a similar preference among storywriters for AI as a collaborative tool rather than a replacement for human creativity.
    Familiar with AI tools like Grammarly\footnote{\url{https://app.grammarly.com/} (last accessed: 09/09/2023)} and DeepL\footnote{\url{https://www.deepl.com/translator} (last accessed: 09/09/2023)}, participants efficiently integrated the LLM into their workflow, accelerating the writing process and amplifying their productivity without sacrificing their sense of accomplishment.
\end{enumerate}

\subsection{Limitations \& Future Work}

Our study focused on young professionals as a proxy for advanced LLM adoption in the knowledge workforce.
Young professionals represent an early majority in LLM adoption, and our participants' demographics represented the current user base of ChatGPT.
However, we emphasize the specific context of our participant group – young professionals in a master's level program – which may limit the generalizability of our findings to other demographics.
Future research could replicate our study with a larger sample, including more experienced workers and a representation of different professions. 

The pre-study interviews and diary study were conducted with a cohort from our university program.
This gave us the unique opportunity for a comparable group of participants regarding their work experience and provided an equivalent set of work tasks.
The tasks used for the diary study (see section \ref{sec:diary_study_setup}) represent the majority of typical knowledge worker tasks \citep{kim_understanding_2019}.
However, not all of these tasks were represented in our diary study.
For example, all participants were physically available at the facilities during the diary study.
This diminished the need for online communication, such as writing emails.
In fact, only one instance in our diary study surveys was related to formulating emails.
Further, our set-up did not include specific knowledge work tasks such as software engineering. Only two participants reported their usage of ChatGPT for code debugging once each. 

Participants in the diary study were asked to use the freely available version of ChatGPT based on GPT-3.5 to ensure comparability across participants. 
With the release of the subscription-based ChatGPT Plus based on GPT-4\footnote{\url{https://openai.com/gpt-4} (last accessed: 09/09/2023)}, some of our findings might not be as prevalent.
For example, the assessed quality of ChatGPT output might have been higher for GPT-4 results than GPT-3.5 results, diminishing the need for post-processing and thus resulting in different accomplishment ratings.
However, as their sense of accomplishment was fundamentally intertwined with their sense of ownership and the sense of ownership being a consistent theme throughout Human-AI automation, we expect our findings to survive time.

Our research depicts the complex assessment of one's perceived productivity and personal accomplishment using LLM tools like ChatGPT. 
Given the only recent mainstream adoption of LLMs and the exploratory nature of our qualitative investigation with its brief timeline and limited sample, the elicited drivers for and barriers to perceived productivity and personal accomplishment serve as a snapshot of current human-AI collaboration in the work context and provide avenues for further research in this rapidly evolving field. 

One of our participants asked, ``If the machine is as good as me, then what use am I?'' hinting at a dystopian future of human-AI interaction.
In our analysis, we discern a complex interplay between the use of GenAI tools like ChatGPT and the professional experiences of young adults. 
While these tools undeniably enhance efficiency and task completion, they also introduce content accuracy and personal accomplishment challenges. 
However, our stance is still optimistic.
Our findings align with a large-scale study by \citet{dellacqua_navigating_2023} conducted on 758 consultants, where AI's impact varies across different tasks, underscoring the importance of a nuanced approach in harnessing AI's capabilities.
Moreover, their study also revealed two effective interaction methods between humans and AI: users with a ``centaur'' approach divided their tasks between the AI and themselves, whereas users with a ``cyborg'' approach fully integrated the AI into their task flow.
In the future, LLM interfaces and interaction methods should be able to cater to these different collaboration styles to effectively enhance one's productivity.

As we specifically looked at young professionals, it was interesting to see how they reflect on being confronted with AI tools and using them in the future. While some see the interaction with AI tools almost as a fun challenge and a skill to add to their toolbox, others are more concerned. 
A recent study by \citet{li_user_2023} indicates that young professionals might be more affected by AI skill degradation and job displacement than senior professionals. 
Ensuring AI literacy, on the one hand, while not neglecting individuals' needs of personal accomplishment in the workplace, on the other hand, should be a research realm worth exploring.
The dual role of AI, as a productivity enhancer and a potential disruptor, calls for a balanced approach in its adoption, emphasizing the importance of user engagement and further Human-Computer Interaction (HCI) research.
\section{Conclusion}

The recent viral adoption of GenAI tools like ChatGPT has the potential to profoundly impact the daily work routines of knowledge workers. This study examined the implications of LLM usage on the self-perception of productivity and accomplishment among young white-collar workers. Our findings reveal that ChatGPT, on the whole, enhances participants' perceptions of productivity and accomplishment, even when it assumes a substantial role in task execution. Key drivers contributing to this heightened sense of accomplishment include the ability to generate greater creative output, the satisfaction derived from the efficient utilization of ChatGPT, and the efficient completion of tasks. 

However, some participants also experienced a decrease in perceived productivity and accomplishment. This decline was primarily attributed to a diminished sense of ownership, a perceived lack of challenge, and the acceptance of mediocre results due to time constraints despite the knowledge that better outcomes would have been attainable. Furthermore, some participants struggled with feelings of inferiority after comparing their outputs with those generated by ChatGPT, which they perceived to be equal or higher quality than their own. 

Moreover, our study suggests that the suitability of task delegation to ChatGPT varies strongly depending on the nature of the task. Participants found employing ChatGPT for research tasks challenging, citing issues such as fabricated sources, necessitating extensive validation. Conversely, users identified ChatGPT as adept at comprehending broad subject domains, generating creative solutions, uncovering new information, and providing creative input, sparing participants from starting from scratch and tackling writer's block.

In this study, we not only highlight the prospects and hurdles of the ongoing transformation in knowledge work but also acknowledge the evolving nature of LLMs and users' expectations and skills. Our findings offer valuable insights into this transitional phase, facilitating the development of best practices and creating user-centric training and innovative interfaces to guide users through this changing landscape.

\bibliographystyle{ACM-Reference-Format}
\bibliography{references}

\clearpage
\appendix
\balance
\section{Pre-Study Interview Script and Guide}
Thank you for joining our pre-study. 
If it is okay for you, we will record this session, transcribe it, and evaluate the results. \textit{<Start recording>}
A remark before we start: You cannot say anything wrong in this interview. This is not a test. \newline

\textbf{Questions}
\begin{enumerate}
    \item How do you use LLMs in the work context?
    \item How do you use LLMs in your university projects?
    \item What UX challenges do you face while using LLMs?
    \item How do LLMs make you feel regarding your productivity?
    \item How does using LLMs make you feel regarding your personal accomplishment? 
\end{enumerate}

Once again, thank you very much for joining our pre-study.

\section{Diary Study Survey} \label{diary_study_survey}
Did you use ChatGPT today? \newline
[ ] Yes \newline
[ ] No

If No:
\begin{enumerate}
    \item What did you mostly work on today? \newline
    \textit{<Free text field>}
    \item Why did you decide not to use ChatGPT today? \newline
    \textit{<Free text field>}
\end{enumerate}

If Yes:
\begin{enumerate}
    \item For which tasks did you use ChatGPT today?\newline
    [ ] Understanding complex or new topics \newline
    [ ] Writing Text \newline
    [ ] Improving Text \newline
    [ ] Research \newline
    [ ] Other:
    \item Did the quality of your work improve by using ChatGPT? \newline
    \textit{<6-point Likert scale from ``strongly disagre'' to ``strongly agre''>}
    \item Why did you feel so? \newline
    \textit{<Free text field>}
    \item Did the quantity of your work improve by using ChatGPT? \newline
    \textit{<6-point Likert scale from ``strongly disagre'' to ``strongly agre''>}
    \item Why did you feel so? \newline
    \textit{<Free text field>}
    \item How accomplished do you feel by your work done with ChatGPT today? \newline
    \textit{<6-point Likert scale from ``not accomplished at all'' to ``very accomplished''>}
    \item Why did you feel so? \newline
    \textit{<Free text field>}
    \item What would improve your experience with ChatGPT? \newline
    \textit{<Free text field>}
\end{enumerate}

\section{Exit Interview Script and Guide} \label{exit_interviews_guide}

Thanks for joining the exit interviews. 
If it’s okay for you, we will record this session, transcribe it, and evaluate the results. \textit{<Start recording>}
Could you please tell me your UID? This information will only be used to map the exit interviews to the results of the diary study.

\begin{itemize}
    \item First two letters of your last name
    \item First two letters of the month you are born in
    \item Last two numbers of your zip code
\end{itemize}

Only the researchers will know the mapping, and we will not share any information.
A remark before we start: You cannot say anything wrong in this interview. This is not a test. Our goal is solely to contextualize the survey findings.

\textbf{Questions}
\begin{enumerate}
    \item How did the usage of ChatGPT change the quality of your output?
    \item How does using ChatGPT affect your sense of accomplishment?
    \item How important is post-processing of ChatGPT output for you?
    \item How does post-processing change your sense of accomplishment?
    \item Imagine you got a perfect output based on your ChatGPT prompt; how would this make you feel?
    \item What were your main motivations for using ChatGPT during the diary study?
    \item How do you now reflect on your ChatGPT usage being part of the diary study?
    \item What else did you find surprising when using ChatGPT and your interaction?
    \item Looking 20 years ahead, what would the ideal interaction between you and ChatGPT look like?
    \item Would you keep using ChatGPT? If so, what would be your primary use cases? Which tasks would you keep for yourself?  
\end{enumerate}

\end{document}